# Design of thermal hysteresis in nonstoichiometric Fe$_2$P-type alloys with giant magnetocaloric effect

Sagar Ghorai [1],[*] Rebecca Clulow,[2] Johan Cedervall,[2] Shuo Huang,[3,4] Tore Ericsson,[2] Lennart Häggström,[2] Ridha Skini,[5] Vitalii Shtender,[2] Levente Vitos,[6] Olle Eriksson,[4,7] Franziska Scheibel,[1] Konstantin Skokov,[1] Oliver Gutfleisch,[1] Martin Sahlberg,[2,7] and Peter Svedlindh[5,8]

[1]*Institute of Material Science, Technical University of Darmstadt, 64287 Darmstadt, Germany*
[2]*Department of Chemistry – Ångström Laboratory, Uppsala University, Box 538, SE-751 21 Uppsala, Sweden*
[3]*Faculty of Materials Science and Chemistry, China University of Geosciences, Wuhan 430074, China*
[4]*Department of Physics and Astronomy, Uppsala University, Box 516, SE-751 20 Uppsala, Sweden*
[5]*Department of Materials Science and Engineering, Uppsala University, Box 35, SE-751 03 Uppsala, Sweden*
[6]*Department of Materials Science and Engineering, Royal Institute of Technology, Stockholm SE-100 44, Sweden*
[7]*Wallenberg Initiative Materials Science for Sustainability, Uppsala University, 75121 Uppsala, Sweden*
[8]*Wallenberg Initiative Materials Science for Sustainability, Uppsala University, 75103 Uppsala, Sweden*

The nonstoichiometric Fe$_2$P-type FeMn$_{(1-x)}$V$_x$(P$_{0.5}$Si$_{0.5}$)$_{1-x}$ alloys ($x = 0, 0.01, 0.02$, and $0.03$) have been investigated as potential candidates for magnetic refrigeration near room temperature. The magnetic ordering temperature decreases with increasing FeV concentration $x$, which can be ascribed to decreased ferromagnetic coupling strength between the magnetic atoms. The strong magnetoelastic coupling in these alloys results in large values of the isothermal entropy change ($\Delta S_M$); 15.7 J/(kg K), at 2 T magnetic field for the $x = 0$ alloy. $\Delta S_M$ decreases with increasing $x$. Results from Mössbauer spectroscopy reveal that the average hyperfine field (in the ferromagnetic state) and average center shift (in the paramagnetic state) have the same decreasing trend as $\Delta S_M$. The thermal hysteresis ($\Delta T_{\text{hyst}}$) of the magnetic phase transition decreases with increasing $x$, while the mechanical stability of the alloys improves due to the reduced lattice volume change across the magnetoelastic phase transition. The adiabatic temperature change $\Delta T_{\text{ad}}$, which highly depends on $\Delta T_{\text{hyst}}$, is 1.7 K at 1.9 T applied field for the $x = 0.02$ alloy.

DOI: 10.1103/PhysRevB.111.224401

## I. INTRODUCTION

The worldwide use of refrigeration systems consumes about 20% of the world's electricity and is responsible for around 7.8% of the global greenhouse emission [1,2]. Solid-state refrigeration, based on the magnetocaloric (MC) effect is a more efficient (20%–30%) and environmentally friendly alternative cooling technology [3]. Although the first near room temperature giant MC material Gd$_5$(Si,Ge)$_4$ was discovered already in 1997, finding a rare-earth free, nontoxic, and economically viable MC material is still a relevent subject for research. Usually, materials with first-order magnetic phase transitions (FMPTs) show a giant MC effect owing to their coupled structural (or lattice) and magnetic degrees of freedom. The FMPT, induced by the coupled magnetic and structural phase changes is generally accompanied by a thermal hysteresis ($\Delta T_{\text{hyst}}$, difference in phase transition temperature while cooling and heating the material). $\Delta T_{\text{hyst}}$ significantly affects important properties of the MC material; i.e., adiabatic temperature change ($\Delta T_{\text{ad}}$), and isothermal entropy change ($-\Delta S_M$) [4]. The most studied MC materials currently include Fe$_2$P-type [5], LaFe$_{13-x}$Si$_x$ [6], NiMn Heusler-type [7], and $MM'X$ ($M$ = Fe or Mn, $M'$ = Co or Ni, $X$ = Si or Ge) [8] based alloys. Apart from $\Delta T_{\text{hyst}}$, another major drawback of most FMPT alloys is the mechanical instability related to the lattice volume change ($\Delta V$) during the phase transition. In this regard, Fe$_2$P-type (Mn,Fe)$_2$(P,Si) alloys with $\Delta V < 1\%$ and with a largely tunable $\Delta T_{\text{hyst}}$ [9] are promising candidates for magnetic refrigeration systems. Although the $\Delta V$ is negligible in the Fe$_2$P-type alloys, the lattice parameter [$\Delta(c/a)$] change, plays a crucial role in determining mechanical stability [10].

In our previous work, we observed that 5 at% of stoichiometric V addition in the metallic site of FeMnP$_{0.5}$Si$_{0.5}$ reduced the magnetic ordering temperature ($T_t$) by 14% and $\Delta T_{\text{hyst}}$ by 61% [11]. It has also been observed that nonstochiometric addition in the metallic site of the (Fe,Mn)$_2$(P,Si) system can influence the MC effect significantly [12]. However, the effect of nonstochiometry in the P/Si site, often caused by P loss during the synthesis process is still unexplored [13]. Moreover, Dung *et al.* [14] have shown that the MC effect in the (Fe,Mn)$_2$(P,Si) system depends on the drastic change of the Fe moment at the magnetic phase transition, changing from weak itinerant magnetism in the paramagnetic (PM) state to a sizable localized magnetic moment in the ferromagnetic (FM) state. The degree of localization of Fe moments in the PM state is expected to increase if the overlap between the 3$d$ and 2$p$ orbitals of Fe and P/Si decreases, which in turn should have an impact on the change of the Fe moment at the magnetic phase transition and hence also on the MC effect. A site-specific Fe-moment fluctuation has also been found in calculations for the pure Fe$_2$P-compound [15]. Therefore, nonstoichiometric addition in the P/Si site may also influence the MC effect by changing the metal to nonmetal ratio in the alloy.

This study focuses on the effect of Fe-moment localization on the MC effect, as well as on reducing $\Delta T_{\text{hyst}}$, in nonstoichiometric (FeMnP$_{0.5}$Si$_{0.5}$)$_{1-x}$(FeV)$_x$ alloys ($x = 0, 0.01, 0.02,$ and $0.03$). In addition, the effect of nonstoichiometry on $\Delta V$, and $\Delta(c/a)$ associated with the magnetoelastic transition and its influence on the mechanical stability of the alloys is investigated.

## II. EXPERIMENTAL DETAILS AND CALCULATION METHOD

Master alloys of FeMnP$_{0.5}$Si$_{0.5}$ and FeV were synthesized by drop synthesis [13] and conventional arc-melting, respectively. Further, stoichiometric amounts of the two master alloys were mixed by hand grinding, pressed into pellets, and vacuum sealed in quartz tubes. The vacuum-sealed alloys were sintered at 1373 K for 1 hr and at 1073 K for 65 hr, followed by quenching in ice water. Four samples were synthesized in this way; (FeMnP$_{0.5}$Si$_{0.5}$)$_{1-x}$(FeV)$_x$, with $x = 0, 0.01, 0.02,$ and $0.03$.

Temperature dependent X-ray powder diffraction (XRPD) data were collected using a Bruker D8 Advance diffractometer with Cu-$K\alpha_1$ radiation, and the data were analyzed using Pawley refinements within the TOPAS6 software [16]. The energy dispersive X-ray spectroscopy (EDX) measurements were performed on a Zeiss Leo 1550 instrument with an Aztec energy dispersive x-ray detector. A constant acceleration spectrometer with $^{57}$CoRh source was used for the collection of Mössbauer spectra at 410 and 100 K. The spectra were folded and fitted using the least square Mössbauer fitting program RECOIL. The magnetic properties were investigated in the temperature range from 5 to 400 K using Quantum Design MPMS-XL and PPMS systems with a maximum magnetic field of 5 T. The adiabatic temperature change ($\Delta T_{\text{ad}}$) was measured in the temperature range from 220 to 380 K, with a maximum applied field of 1.9 T in a home-built device at the Technical University of Darmstadt, Germany [17].

Density functional theory [18] calculations were carried out by the exact muffin-tin orbitals method [19]. The one-electron Kohn-Sham equations were solved within the soft-core and scalar-relativistic approximations. The self-consistent calculations were performed with the local-density approximation (LDA) by Perdew and Wang [20], and the total energy calculations were performed with the generalized-gradient approximation by Perdew, Burke, and Ernzerhof (PBE) [21]. The PBE correction is applied non-self-consistently on the charge density obtained from the LDA calculations to enable a balance between computational efficiency and accuracy.

The Green's function was calculated by using 16 complex energy points on a semicircular contour including the valence states. The chemical disorder was treated by the coherent-potential approximation [22]. The PM state was simulated by the disordered local moments approximation [23]. Further details about the adopted methods can be found in previous work [19].

## III. PHASE STABILITY

Total energy calculations for the studied alloys were performed to check the phase stability and the site preference of the atoms. From previous studies [11,24], it is known that Fe, Mn, and P/Si prefer to occupy the 3$f$, 3$g$, and 1$b$/2$c$ crystallographic sites in the hexagonal Fe$_2$P-type structure, respectively [cf. Fig. 1(a)]. For the total energy calculation, four possible cases have been considered: (i) V occupies only the 3$f$ site ($V_{3f}$); (ii) V occupies only the 3$g$ site ($V_{3g}$); (iii) Fe, Mn, and V randomly occupy the 3$f$ and 3$g$ sites ($A_{\text{random}}$); and (iv) Fe and Mn from the Fe$_2$P-type phase occupy the 3$f$ and 3$g$ sites, respectively, while Fe and V from the FeV phase occupy the 3$f$ and 3$g$ sites randomly ($B_{\text{random}}$). The total energy values as a function of lattice volume for all cases in the FM state are shown in Figs. 1(b)–1(d). Owing to the relatively higher total energy values, the $A_{\text{random}}$-case has been excluded from Figs. 1(b)–1(d). In addition, for better comparison between different cases, the total energy values are shifted by the minimum energy [see insets of Figs. 1(b)–1(d)] of each corresponding alloy. The minimum energy is observed for the $V_{3g}$ case, i.e., when V occupies only the 3$g$ site. Noticeably, the relative energy difference increases with increasing $x$, confirming a more stable V occupancy in the 3$g$ site. A similar 3$g$ site preference of V is also calculated for the PM state [25]. In addition to the total energy calculation, the formation energy calculation of the (FeMnP$_{0.5}$Si$_{0.5}$)$_{1-x}$(FeV)$_x$ alloy, indicates a 3$g$-site preference of V, which is discussed in the SI [25]. Therefore, considering the V occupancy in the 3$g$ site, the chemical formula for the studied alloys can be written as, FeMn$_{(1-x)}$V$_x$(P$_{0.5}$Si$_{0.5}$)$_{1-x}$.

## IV. RESULTS AND DISCUSSION

### A. Chemical composition analysis

The chemical compositions and homogeneity of the alloys have been verified by EDX analysis. Owing to the porous and brittle nature of the alloys, the possibility to resolve proper phase boundaries is limited. However, from the elemental mapping of the studied alloys, apart from the main Fe$_2$P-type phase, a small amount of secondary phase with P and Mn deficiency has been observed [see highlighted regions in Figs. 2(a)–2(d)]. The elemental at% of the main Fe$_2$P-type phase is shown in Fig. 2(e). The variation of V in the alloys is as expected, while the amount of Fe in the main phase is lower than expected, indicating that Fe is involved in the formation of the secondary phase. The observed chemical composition of the secondary phase is; Fe 44(4) at%, Mn 30(3) at%, P 8.3(8) at% and Si 18(2) at% for the $x = 0$ alloy. A similar P-deficient secondary phase is also observed for the rest of the alloys. From XRPD analysis (will be discussed later), this secondary phase is identified as a (Fe,Mn)$_3$Si-type phase which is also observed for stoichiometric V added alloys [11].

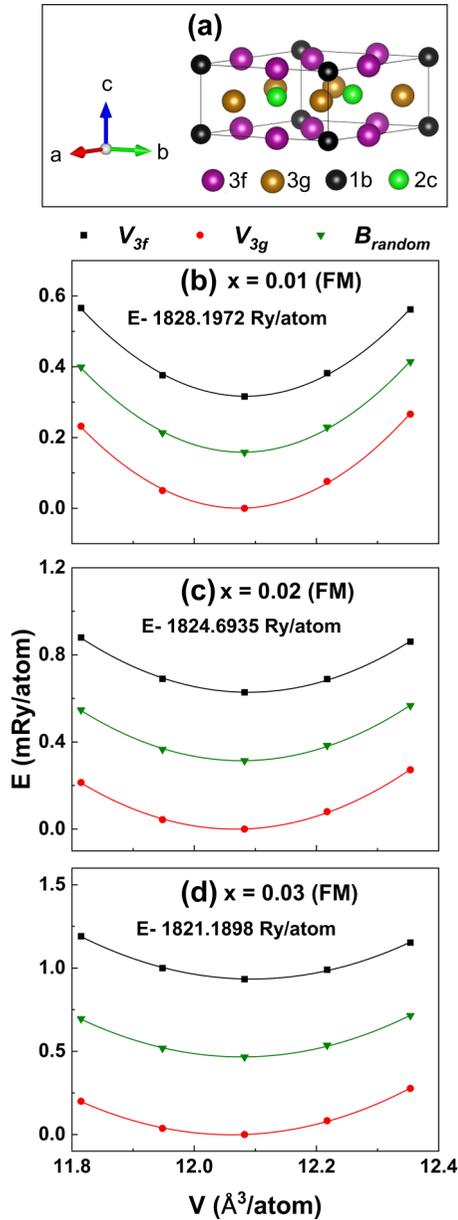

FIG. 1. (a) Hexagonal structure of the FeMnP$_{0.5}$Si$_{0.5}$ alloy obtained from XRPD refinement; the different crystallographic sites are color coded. [(b)–(d)] The calculated total energy as a function of the lattice volume in the FM state of the studied alloys. The energy values are rescaled (see inset formula) with respect to the minimum total energy of the corresponding alloy. The color coded lines joining the data points are polynomial fits to the corresponding data. See text for the notations of different configurations in the legend.

### B. Coupled structural and magnetic phase transition

The analysis of the temperature dependent XRPD data reveals a hexagonal Fe$_2$P-type structure with space group $P\bar{6}2m$ for the studied alloys [25]. Although the hexagonal structure remains unchanged, the $c/a$ ratio exhibits a strong change with temperature in the magnetic transition region (to be discussed later), which is often referred to as a magnetoelastic phase transition. With decreasing temperature, the $c/a$ ratio changes from a high ($\approx 0.57$) to a low ($\approx 0.53$) value [see Figs. 3(a) and 3(b)]. During this lattice parameter change, there is also a change of lattice volume ($\Delta V$), as shown in the inset of Fig. 3(a). $\Delta V$ is defined as the volume difference between the low and high $c/a$-ratio phases. The value of $\Delta V$ decreases with increasing $x$, except for the $x = 0.03$ alloy. The lattice parameter change is accompanied by a magnetic phase transition [PM to FM phase, cf. Fig. 3(c)], enabling coupled magnetic and lattice entropy changes. The sharpness of the $c/a$-ratio change across the transition region indicates the strength of the magnetoelastic phase transition, which can be defined as [11]

$$\Delta(c/a)(\%) = \lim_{T \to T_t} \frac{(c/a)_{PM} - (c/a)_{FM}}{(c/a)_{PM}} \times 100, \quad (1)$$

where $(c/a)_{PM}$ and $(c/a)_{FM}$ correspond to the $(c/a)$-ratio just above and below the magnetic ordering temperature $T_t$, respectively. From the inset of Fig. 3(b), the largest (smallest) change of the $(c/a)$ ratio has been observed for the $x = 0.03$ ($x = 0.01$) alloy.

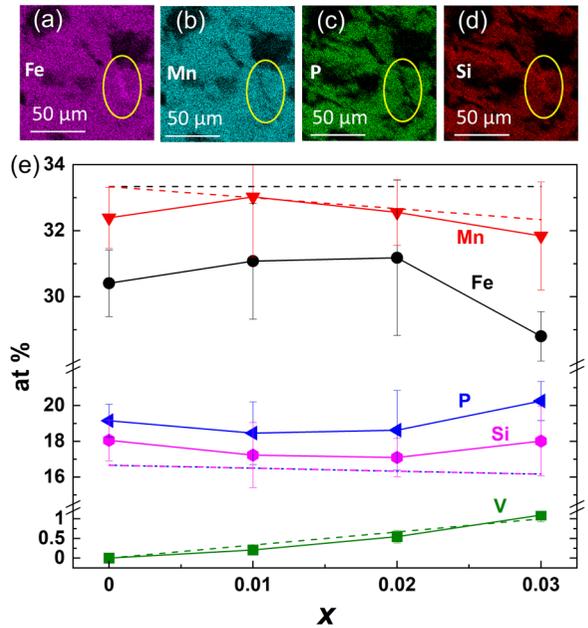

FIG. 2. [(a)–(d)] Elemental mapping of the $x = 0$ alloy. The area marked with a yellow ellipse represents the secondary phase. A brighter (darker) color indicates an increased (decreased) amount of the element. (e) Variation of elemental at% with $x$. The dotted lines with same color indicate the desired at% of the element in the FeMn$_{(1-x)}$V$_x$(P$_{0.5}$Si$_{0.5}$)$_{1-x}$ alloys. The error bars correspond to standard deviations of ten successive measurements.

The value of $T_t$ decreases as expected with increasing V addition in the metal site of the alloy [cf. Fig. 3(c)]. However, one should keep in mind that the nonstoichiometry in the nonmetal site also influences $T_t$. As a comparison, with 5 at% stochiometric V addition, $T_t$ decreased by 14%, while in the present study, with 3 at% nonstoichiometric V addition $T_t$ decreases by 25% [11]. This additional decrease of $T_t$ is the result of the nonstoichiometry in the nonmetal site. The XRPD analysis reveals that apart from the principle Fe$_2$P-type hexagonal phase all samples contain the (Fe,Mn)$_3$Si-type secondary phase. However, the parent alloy additionally contains a small amount (3 wt%) of a Mn$_5$Si$_3$-type secondary phase, which is negligible in the FeV added alloys [25]. With increasing $x$, there is an almost linear increase of the (Fe,Mn)$_3$Si phase (from 3.1 wt% in the $x = 0$ alloy to 9.2 wt% in the $x = 0.03$ alloy). This secondary phase is magnetic, the small kinks observed in the temperature dependent magnetization curves [cf. Fig. 3(c)] below $T_t$ originate from this secondary phase [11,26]. The anomalous $\Delta V$ and $\Delta(c/a)$ values for the $x = 0.03$ alloy [cf. insets of Figs. 3(a) and 3(b)] is mostly the result of the larger weight-% of the secondary phase. The same secondary phase has also been observed for stoichiometrically V added FeMnP$_{0.5}$Si$_{0.5}$ alloys. However, with stochiometric V addition, the amount of the secondary phase was almost constant [11]. Irrespective of the metal to nonmetal ratio, the presence of the secondary phase will decrease the Si/P ratio in the alloy. Previously, a decrease of $T_t$ has been observed with decreasing Si/P ratio [14,27], without providing an explanation for the Si/P ratio dependence of $T_t$.

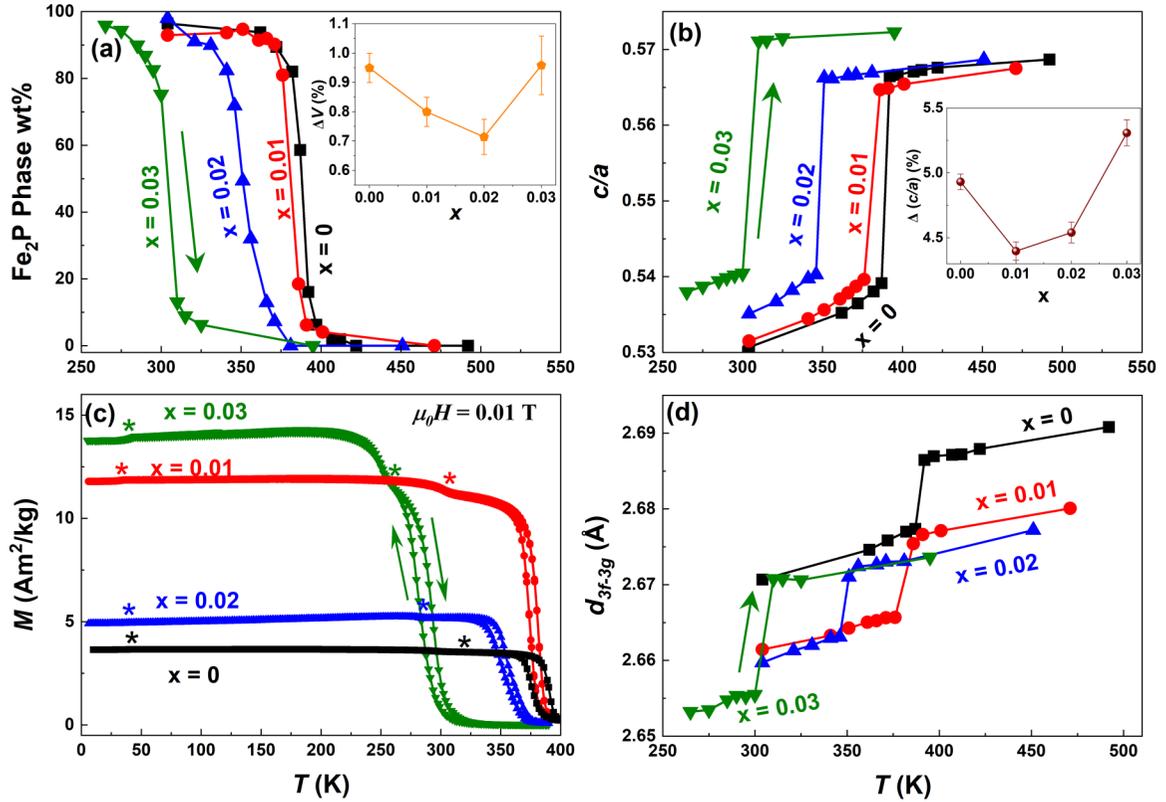

FIG. 3. (a) Temperature dependence of the low $c/a$ ratio ($\approx 0.53$) Fe$_2$P phase, data recorded during heating; the inset shows the change in lattice volume $\Delta V$ in the transition region versus $x$, being defined as the difference in lattice volume between the low and high $c/a$-ratio phases. (b) $c/a$-ratio versus temperature considering only the dominating Fe$_2$P-phase (wt% >60%) during heating. The calculated error from the XRPD refinement of the lattice parameter data is in the order of $10^{-4}$ Å. The inset shows the change of the $c/a$ ratio vs $x$ in the magnetic transition region. (c) Magnetization versus temperature at $\mu_0 H = 0.01$ T; the kinks in the magnetization curves (marked with $*$) originate from the magnetic secondary phase. (d) Atomic distance between the $3f$ and $3g$ sites versus temperature.

It has been observed that for different atomic substitutions in the metallic site, $T_t$ has increased or decreased in the Fe$_2$P system [28,29]. Considering only metallic bonds in the Fe$_2$P-system, $T_t$ should be related to the direct exchange interaction ($J$) between magnetic atoms. The Bethe-Slatter (BS) curve provides a relationship between $J$ and the nearest-neighbor distance (normalized by the atomic radius, $d/a$) [30]. From the BS curve, in the ferromagnetic region, $J$ increases up to a certain threshold value of $d/a$, after which it starts to decrease. Therefore, it can be concluded that $T_t$ will also increase up to a certain threshold value of the atomic distance, after which it will decrease. For the Fe$_2$P system, $d/a$ can be related to the atomic distance between the magnetic atoms situated in the $3f$ and $3g$ sites. From the XRPD data refinement, the positions of the $3f$ and $3g$ sites atoms (cf. Fig. 1) are $(x, 0, 0)$, and $(y, 0, 0.5)$, respectively. Using these values, the interatomic distance ($d_{3f-3g}$) between the atoms in the $3f$- and $3g$-sites is calculated as [31]

$$d_{3f-3g} = \sqrt{[(y-x) \times a]^2 + [0.5 \times c]^2}. \quad (2)$$

The values of $d_{3f-3g}$ are shown in the Fig. 3(d). It is clear that the values of $d_{3f-3g}$ decrease with increasing $x$ following a similar trend as $T_t$.

The saturation magnetizations ($M_S$) for the studied alloys are estimated from the magnetic field dependent magnetization measurement at 5 K temperature (cf. Fig. 4). With increasing $x$, the decrease of $M_S$, results from the replacement of magnetic atoms (Mn or Fe) with nonmagnetic V atoms. Interestingly, the magnetization at lower fields does not follow the trend of the $M_S$, and indicates a change in magnetic anisotropy or demagnetization factor with $x$.

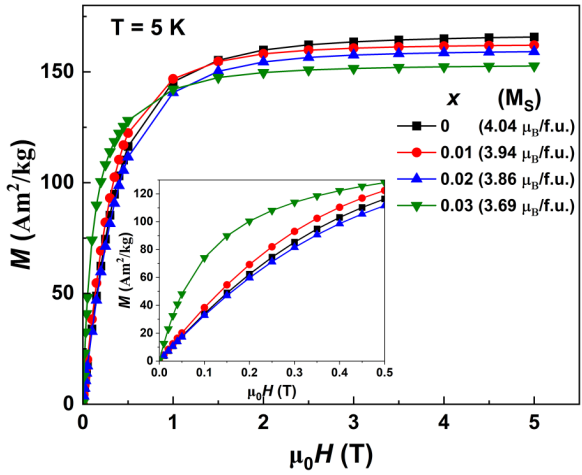

FIG. 4. Magnetic field dependent magnetization at 5 K temperature. The inset highlights the low-field region.

### C. Magnetocaloric effect and thermal hysteresis

The MC effect can be quantified by the isothermal magnetic entropy change ($\Delta S_M$) of the system. For a magnetic field change from zero to $H_f$, $\Delta S_M$ can be expressed using Maxwell's relation as [32]

$$\Delta S_M(T, H_f) = -\mu_0 \int_0^{H_f} \left(\frac{\partial M}{\partial T}\right)_H dH. \quad (3)$$

The calculated values of $\Delta S_M$ for the studied alloys are shown in Fig. 5(a). A decrease of $\Delta S_M$ with increasing $x$ is observed. In our previous work [11], for the stoichiometrically V added FeMnP$_{0.5}$Si$_{0.5}$ alloys, it was observed that the value of $\Delta S_M$ is directly proportional to the strength of the magnetoelastic coupling. The strength of the magnetoelastic coupling is defined by the relative change of the hexagonal lattice parameters. In this work, the relative change of the $c/a$ ratio at the magnetic transition is highest for the $x = 0.03$ alloy [cf. inset of Fig. 3(b)]. However, the trend of $\Delta S_M$ does not follow the lattice parameter change. Therefore the magnetoelastic coupling strength can not be described from the lattice parameters of the nonstoichiometric Fe$_2$P-type alloys, especially when there is a magnetic secondary phase involved.

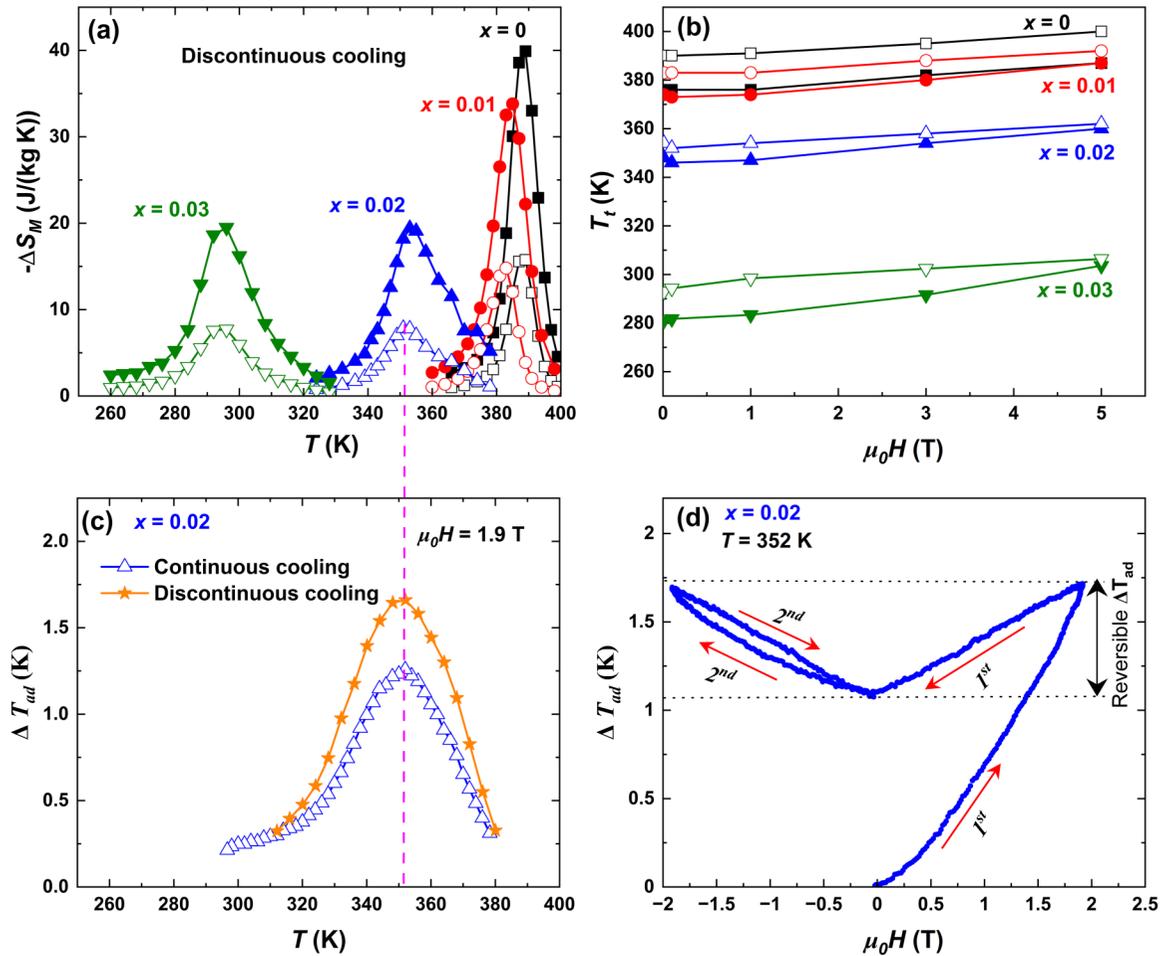

FIG. 5. (a) Temperature dependent isothermal entropy change measured at 2 T (hollow symbols) and 5 T (solid symbols) applied fields, respectively. (b) $T_t$ vs magnetic field, solid (hollow) symbols correspond to data recorded during cooling (heating). (c) Temperature dependent adiabatic temperature change for the $x = 0.02$ alloy, measured at $\mu_0H = 1.9$ T following continuous cooling (hollow symbols) and discontinuous cooling (solid symbols) protocols. (d) Magnetic field dependent irreversibility of $\Delta T_{ad}$ at 352 K. Arrows indicate the direction of magnetic field change.

TABLE I. Magnetocaloric properties of the studied alloys (*) compared with data reported for other giant MC materials near room temperature. The $\Delta S_M$ values correspond to a $\mu_0 H = 2$T field change. The values of $T_t$ for the studied alloys are taken from the field cooled (FC) magnetization versus temperature curves at $\mu_0 H = 0.01$ T.

| Sample | $T_t$ (K) | $-\Delta S_M$ (J/kg) | $\Delta T_{\mathrm{ad}}$ ($\mu_0 H$) (K) | Ref. |
|---|---|---|---|---|
| FeMnP$_{0.5}$Si$_{0.5}$ | 376 | 16.5 | – | * |
| FeMn$_{0.99}$V$_{0.01}$(P$_{0.5}$Si$_{0.5}$)$_{0.99}$ | 374 | 14.8 | – | * |
| FeMn$_{0.98}$V$_{0.02}$(P$_{0.5}$Si$_{0.5}$)$_{0.98}$ | 348 | 7.8 | 1.7 (1.9 T) | * |
| FeMn$_{0.97}$V$_{0.03}$(P$_{0.5}$Si$_{0.5}$)$_{0.97}$ | 281 | 7.7 | – | * |
| FeMn$_{0.95}$V$_{0.05}$P$_{0.5}$Si$_{0.5}$ | 322 | 13.1 | – | [11] |
| Fe$_{0.95}$V$_{0.05}$MnP$_{0.5}$Si$_{0.5}$ | 318 | 9.1 | – | [11] |
| Fe$_{0.71}$Mn$_{1.32}$P$_{0.5}$Si$_{0.56}$ | 265 | 16 | 2.35 (1.9 T) | [35] |
| Fe$_{0.975}$Mn$_{0.975}$P$_{0.47}$Si$_{0.5}$B$_{0.03}$ | 329 | 11.6 | 1.9 (2 T) | [36] |
| Fe$_{0.84}$Co$_{0.11}$MnP$_{0.51}$Si$_{0.45}$B$_{0.04}$ | 295 | 11.4 | 1.9 (1.1 T) | [37] |
| FeMnP$_{0.45}$As$_{0.55}$ | 306 | 10.7 | 2.9 (1.1 T) | [38] |
| Ni$_{45.2}$Mn$_{36.7}$In$_{13}$Co$_{5.1}$ | 311 | ∼19 | 6.2 (1.9 T) 1.5 (−1.9 T) | [17] |
| La$_{0.4}$Pr$_{0.3}$Ca$_{0.1}$Sr$_{0.2}$MnO$_3$ | 289 | 3.08 | 1.5 (1.9 T) | [39] |
| Gd | 295 | 6.1 | 5.5 (1.9 T) | [40] |

Since the total entropy during an adiabatic process is conserved, the change of magnetic entropy by the application or removal of a magnetic field will change the phonon entropy of the system. This change of phonon entropy will also change the temperature of the system, known as the adiabatic temperature change ($\Delta T_{\mathrm{ad}}$), the most important parameter for characterization of the MC effect [33]. For a system with a heat capacity of $C(H, T)$, $\Delta T_{\mathrm{ad}}$ for a magnetic field change from zero to $H_f$ can be expressed as [34]

$$\Delta T_{\mathrm{ad}}(T, H_f) = -\mu_0 \int_0^{H_f} \left(\frac{T}{C}\right)_H \frac{\partial M}{\partial T}\bigg|_H dH. \quad (4)$$

From Eq. (3), it is clear that at a particular temperature, $-\Delta S_M$ depends both on the magnitude of the magnetization and rate of change of the magnetization with respect to temperature. In addition to this, Eq. (4) indicates that a smaller heat capacity yields a larger $\Delta T_{\mathrm{ad}}$. In this work, the $\Delta T_{\mathrm{ad}}$ has only been measured for the $x = 0.02$ alloy, as the rest of the alloys were extremely brittle, implying that $\Delta T_{\mathrm{ad}}$ of these samples could not be measured with our setup where physical contact between the sample and the thermocouple is required. This mechanical instability of the three alloys can also be related to the relatively large value of $\Delta V$ [cf. inset of Fig. 3(a)]. The MC effect properties of the studied alloys are compared in Table I with well-known giant MC materials with magnetic phase transitions near room temperature.

The observed values of $\Delta T_{\mathrm{ad}}$ for the $x = 0.02$ alloy, measured following continuous cooling and discontinuous cooling (where the sample is heated to its PM state before approaching the measuring temperature) protocols are shown in Fig. 5(c). The large difference between measured $\Delta T_{\mathrm{ad}}$ values, following the continuous cooling and discontinuous cooling protocols, is due to the thermal hysteresis $\Delta T_{\mathrm{hyst}}$ of the alloy [4]. To understand the effect of $\Delta T_{\mathrm{hyst}}$, the field dependence of $\Delta T_{\mathrm{ad}}$ is shown at a temperature (352 K) near the $T_t$ of the alloy. Before this measurement, the sample was heated to its PM state, i.e., $\Delta T_{\mathrm{ad}}$ was recorded following the discontinuous protocol. While cooling from the PM state, in the vicinity of $T_t$, the sample will undergo a magnetic phase transition coupled with a structural (in this case a $c/a$-ratio change) phase transition. As shown in Fig. 5(d), starting from zero magnetic field and increasing the field to 1.9 T the temperature of the sample increases by 1.7 K, but removal of the field does not bring the sample back to its initial temperature. However, applying and removing the same field (this time −1.9 T) a second time will bring the sample back to the same temperature as obtained after the first field cycle. In this sense $\Delta T_{\mathrm{ad}}$ is reversible during the second field cycle. The irreversible behavior of $\Delta T_{\mathrm{ad}}$ is a result of the coupled magnetic and structural phase changes; as the structural phase change is irreversible, $\Delta T_{\mathrm{ad}}$ will also be irreversible during the first magnetic field cycle [41]. Mostly, the residual heat after the first cycle represents the latent heat for the structural phase change. During the second magnetic field cycle, the applied field can not overcome $\Delta T_{\mathrm{hyst}}$, as a result, the magnetoelastic phase transition is not complete. Similar results have been observed for the Ni-Mn-In-Co Heusler compound [42].

The field dependence of $T_t$ during heating and cooling is shown in Fig. 5(b). The temperature region between the heating and cooling curves corresponds to a mixed paramagnetic-ferromagnetic state defining $\Delta T_{\mathrm{hyst}}$. The lowest value of $\Delta T_{\mathrm{hyst}}$ is observed for the $x = 0.02$ alloy. To predict the tricritical point where $\Delta T_{\mathrm{hys}} = 0$, a linear extrapolation of the data in Fig. 5(b) yields the critical temperature ($T_{\mathrm{crit}}$) and critical field ($\mu_0 H_{\mathrm{crit}}$) for each alloy, except for the $x = 0$ alloy for which the used magnetic field range was not sufficient to make this analysis meaningful. A $\mu_0 H_{\mathrm{crit}}$ value of 10 T is obtained for the $x = 0.01$ alloy, while $\mu_0 H_{\mathrm{crit}} \approx 6$ T is obtained for the $x = 0.02$ and $x = 0.03$ alloys, with the corresponding $T_{\mathrm{crit}}$ values being 404, 365, and 309 K, respectively.

### D. Hyperfine interaction

From the previous discussion, it is clear that for nonstoichiometric Fe$_2$P-type alloys the magnetoelastic coupling strength, derived from the lattice parameter change at $T_t$, can not properly describe the difference in $\Delta S_M$ between the alloys. The large $\Delta S_M$ values of the (Fe,Mn)$_2$(P,Si) alloys are often ascribed to the Fe-moment fluctuation at $T_t$ and its hybridization with the nonmetallic atoms in the alloy [14]. Mössbauer spectra in the PM (410 K) and FM (100 K) states of the studied alloys have been collected to investigate the hyperfine interaction of Fe with its neighboring atoms; the results are presented in Figs. 6(a) and 6(b), respectively.

From the spectra recorded at 410 K, the strong $3f$ site preference for Fe is confirmed by the absence of any secondary line at higher velocity [43]. All 410 K spectra irrespective of V content were fitted with three doublets corresponding to three possible nearest neighbor interactions of the Fe atom. These are one Fe atom surrounded by two P and two Si atoms (Fe$_{2P2Si}$), three P and one Si atom (Fe$_{3P1Si}$), and four

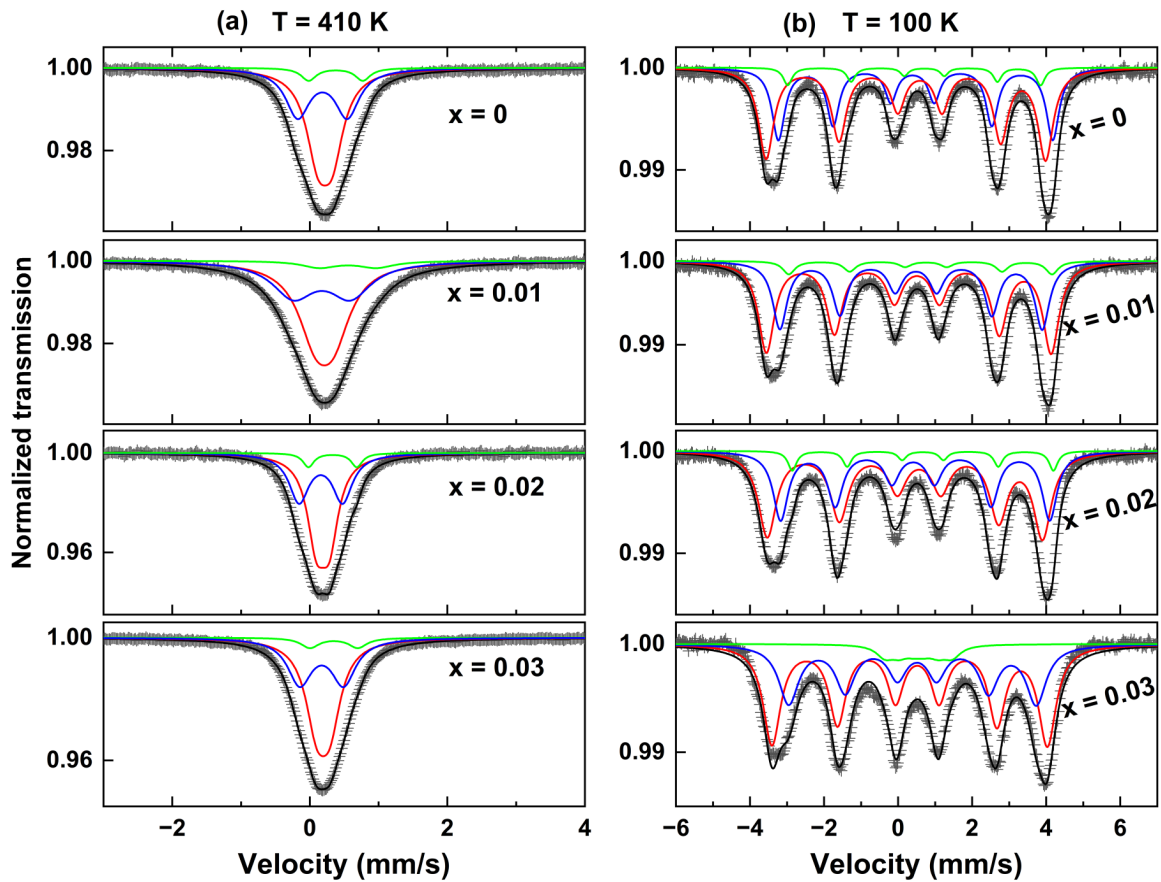

FIG. 6. Mössbauer spectra at (a) 410 and (b) 100 K. The red, blue, and green subpatterns correspond to the nearest neighbor surroundings, Fe$_{2P2Si}$, Fe$_{3P1Si}$, and Fe$_{4P}$ of Fe at the 3$f$ site, respectively.

P atoms (Fe$_{4P}$) with probabilities 0.5625, 0.375, and 0.0625, respectively. The probabilities are calculated using the site preference for the 2$c$ site for Si [24]. To shed light on the Fe-moment localization and its variation with $x$, the average center shift (*CS*) at 410 K and the average hyperfine field ($B_{hf}$) at 100 K are presented in Fig. 7. The reported CS values have natural $\alpha$-Fe at 295 K as a reference absorber. The average *CS* value decreases with increasing $x$. This decrease corresponds to an enhanced electron density at the Fe nuclei, which indicates a stronger orbital overlap between the 3$d$ and 2$p$ orbitals of the Fe and P/Si atoms. As mentioned in the introduction, the large value of $\Delta S_M$ in the (Fe,Mn)$_2$(P,Si) alloys can be ascribed to the drastic change of the Fe moment (from $\approx 0.003 \mu_B$/atom to 1.54$\mu_B$/atom) at $T_t$ [14], something which is consistent with results from theoretical calculations [15]. From the observed values of *CS*, the tendency for localization of the Fe moments decreases with increasing $x$, which suggests a suppressed change of the Fe moment at the magnetic transition. With increasing $x$ (cf. Fig. 7), the gradual decrease of $B_{hf}$ in the FM region indicates a decrease of the local magnetization of Fe, which is also in agreement with the decrease of $\Delta S_M$ [43].

It is important to keep in mind that the amount of secondary phase increases with increasing $x$. However, the effect of the secondary phase on the Mössbauer spectra is not obvious.

## V. CONCLUSIONS

The magnetic and magnetocaloric properties of FeMn$_{(1-x)}$V$_x$(P$_{0.5}$Si$_{0.5}$)$_{1-x}$ alloys, with $x = 0$, 0.01, 0.02, and 0.03, have been investigated. From the formation energy calculations, it was found that V has a preferred 3$g$-site occupancy in the Fe$_2$P type hexagonal structure. The studied alloys exhibit temperature dependent magnetoelastic phase transitions, from the low-temperature hexagonal phase with a low $c/a$ ratio to the high-temperature hexagonal phase with a high $c/a$ ratio. Similar to the nonstoichiometry in the metallic site of (Fe,Mn)$_2$(P,Si) alloys, the nonstoichiometry in the nonmetallic site also influences the secondary (Fe,Mn)$_3$Si-type phase formation [12]. A decrease of the phase transition temperature with increasing $x$ is observed, which is an effect of decreased FM exchange interaction between the magnetic atoms (Fe and Mn). The $3f$-site preference of the Fe-atom has been confirmed from analysis of the Mössbauer spectra. In addition, the analysis of the Mössbauer spectra has revealed an enhanced electron density at the Fe nuclei with increasing $x$. Moreover, with increasing $x$, the $B_{hf}$ and the saturation magnetization decreases gradually, providing an explanation for the decrease of $\Delta S_M$.

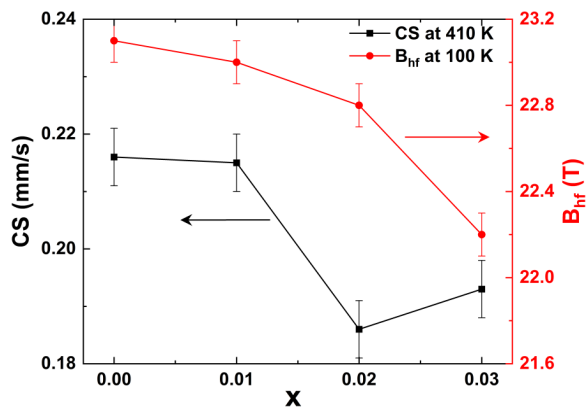

FIG. 7. Average *CS* and $B_{hf}$ values obtained from Mössbauer spectra at 410 and 100 K, respectively. For details see main text.

For stochiometric V addition in the FeMnP$_{0.5}$Si$_{0.5}$ alloys, the variation of $\Delta S_M$ can be directly related to the variation of the magnetoelastic coupling strength, which is given by the relative change of the hexagonal lattice parameters at $T_t$ [11]. However, in the case of nonstochiometric V addition, a secondary phase will form in the alloy. Thus, the magnetoelastic coupling strength can not be directly related to the relative change of lattice parameters in the studied alloys. A magnetic field dependent irreversible variation of the $\Delta T_{ad}$ has been observed for the $x = 0.02$ alloy. The associated $\Delta T_{hyst}$, has been identified as the reason for this irreversibility. Moreover, with nonstoichiometric V addition, $T_t$ can be tuned towards room temperature; $\Delta T_{hyst}$, and $\Delta V$ can be reduced while maintaining a reasonably large value of $\Delta S_M$ and a moderate value of $\Delta T_{ad}$, making this study important for magnetic refrigeration applications and sustainable energy solutions.

## ACKNOWLEDGMENTS

The authors thank the Swedish Foundation for Strategic Research (SSF), project "Magnetic materials for green energy technology" (contract EM-16-0039) for financial support. The authors acknowledge support from STandUPP and eSSENCE. The computational studies were performed on resources provided by the National Academic Infrastructure for Supercomputing in Sweden(NAISS). S.G., F.S., O.G., and M.S. thankfully acknowledge the financial support of the German Research Foundation (DFG) in the framework of the Collaborative Research Centre Transregio 270 (CRC-TRR 270) and the European Union, through project CoCoMag (Grant No. 101099736). J.C. acknowledges support from ÅForsk Foundation (Grant No. 22-378). O.E., P.S., and M.S. acknowledge support from WISE, Wallenberg Initiative Materials Science.